\documentclass[%
 reprint,
 amsmath,amssymb,
 aps,
 superscriptaddress,
prl,
showkeys,
]{revtex4-2}

\usepackage{graphicx}% Include figure files
\usepackage{dcolumn}% Align table columns on decimal point
\usepackage{bm}% bold math
\usepackage{hyperref}% add hypertext capabilities
\hypersetup{colorlinks, allcolors=blue}
\usepackage{physics}
\usepackage[version=4]{mhchem}
\usepackage{siunitx}
\usepackage{lipsum}
\usepackage[utf8]{inputenc}
\usepackage{soul} %for highlighting text

\begin{document}

% Title
\title{Probing the Atomic Arrangement of Sub-Surface Dopants in a Silicon Quantum Device Platform}%

% Authors
\author{H{\aa}kon I. R{\o}st}
\affiliation{Center for Quantum Spintronics, Department of Physics, Norwegian University of Science and Technology (NTNU), NO-7491 Trondheim, Norway.}
\affiliation{Department of Physics and Technology, University of Bergen (UiB), All\'egaten 55, 5007 Bergen, Norway.}

\author{Ezequiel Tosi}
\affiliation{Elettra - Sincrotrone Trieste, s.s. 14 - km.163,5 in Area Science Park, Basovizza, Trieste, 34149, Italy.}
\affiliation{Instituto de Ciencia de Materiales de Madrid (ICMM-CSIC) C/ Sor Juana Inés de la Cruz 3, 28049, Madrid, Spain.}

\author{Frode S. Strand}
\author{Anna Cecilie {\AA}sland}
\affiliation{Center for Quantum Spintronics, Department of Physics, Norwegian University of Science and Technology (NTNU), NO-7491 Trondheim, Norway.}

\author{Paolo Lacovig}
\author{Silvano Lizzit}
\affiliation{Elettra - Sincrotrone Trieste, s.s. 14 - km.163,5 in Area Science Park, Basovizza, Trieste, 34149, Italy.}

\author{Justin W. Wells}
\email[Corresponding author: ]{j.w.wells@fys.uio.no}
\affiliation{Center for Quantum Spintronics, Department of Physics, Norwegian University of Science and Technology (NTNU), NO-7491 Trondheim, Norway.}
\affiliation{Centre for Materials Science and Nanotechnology, University of Oslo (UiO), Oslo 0318, Norway.}

\begin{abstract}
High-density structures of sub-surface phosphorus dopants in silicon continue to garner interest as a silicon-based quantum computer platform, however, a much-needed confirmation of their dopant arrangement has been lacking. In this work, we take advantage of the chemical specificity of X-ray photoelectron diffraction to obtain the precise structural configuration of \ce{P} dopants in sub-surface Si:P $\delta$-layers. The growth of $\delta$-layer systems with different levels of doping is carefully studied and verified using X-ray photoelectron spectroscopy and low-energy electron diffraction. Subsequent XPD measurements reveal that in all cases, the dopants primarily substitute with Si atoms from the host material. Furthermore, no signs of free carrier-inhibiting P$-$P dimerization can be observed. Our observations not only settle a nearly decade-long debate about the dopant arrangement but also demonstrate that XPD is well suited to study sub-surface dopant structures. This work thus provides valuable input for an updated understanding of the behavior of Si:P $\delta$-layers and the modeling of their derived quantum devices.
\end{abstract}

\keywords{quantum devices, delta-layers, quantum computing, photoelectron diffraction}%Use showkeys class option if keyword
                              %display desired
\maketitle
% ===============================================
% THE PAPER BEGINS HERE
% ===============================================
\message{columnwidth = \the\columnwidth and textwidth = \the\textwidth}

\section*{Introduction}
Over the last decade, the effort to realize a silicon-based, CMOS-compatible quantum computer has been intensifying \cite{zwanenburg2013silicon,Veldhorst:2017,Gonzalez-Zalba:2021}, and several significant breakthroughs have been achieved \cite{Sigillito:2019,Ciriano-Tejel:2021,Pauka:2021}.  One common factor in this development is the so-called Si:P $\delta$-layer platform \cite{fuechsle2012single,veldhorst2015two}; i.e. an ultra-sharp and narrow layer of phosphorus dopants placed beneath the silicon surface, which can be patterned with atomic precision \cite{Schofield:2003}. The $\delta$-layer platform can be used for quantum dots and tunnel barriers \cite{Fuechsle:2010}, metallic interconnects \cite{Weber:2012}, and other key components required for quantum device engineering \cite{zwanenburg2013silicon}. This, in turn, has required it to be thoroughly studied and understood \cite{Goh:2004,goh2009impact,Drumm:2012,McKibbin2010investigating,Polley:2012,Miwa2013direct,miwa2014valley,mazzola2014disentangling,Holt2020observation,mazzola2020sub}. Despite these intense efforts and the great progress which has been made, one key question remains unanswered: \emph{What is the arrangement of the dopants within the $\delta$-layer?} The answer is of central importance for the performance of $\delta$-layer-derived devices, because the dopant arrangement is understood to directly impact key electronic properties; for example, the energy separation (i.e. `valley-splitting') of the supported quantum well states  \cite{Carter2009electronic,Lee2011electronic,Carter2011phosphorous}.

There may be multiple reasons why the atomic arrangement is not known, but we conjecture that it is primarily because, until now, a suitable probing method had not been identified. Traditional X-ray diffraction methods are unsuitable because of the atomically thin nature of the $\delta$-layer \cite{chubarov2018plane}. High-angle annular dark-field imaging with an electron microscope is also exceptionally challenging, because of the similarity in atomic weight of Si and P \cite{yamashita2018atomic}. Recent studies have shown that the quantum confinement of the $\delta$-layer can be ascertained by means of ellipsometry \cite{Sandia}, but the in-plane coordination of the dopants has remained elusive.

In this work, we demonstrate that the neighborhood around the dopants can be directly probed using X-ray photoelectron diffraction (XPD), in which a chemically specific diffractive image is formed by utilizing subtle core level energy shifts that are concomitant with the coordination of a dopant \cite{hufner2013photoelectron,Bengio2007structure}. Although XPD is primarily used as a probe of surface structure \cite{Woodruff2007adsorbate,Bignardi2019growth,holt2021electronic}, we demonstrate here that it also has great potential for determining the local arrangement of sub-surface atoms and, therefore, is perfectly suited for solving the long-standing mystery of the Si:P $\delta$-layer structure.

\section*{Results and Discussion}

\begin{figure*}[t]
    \centering
    \includegraphics[]{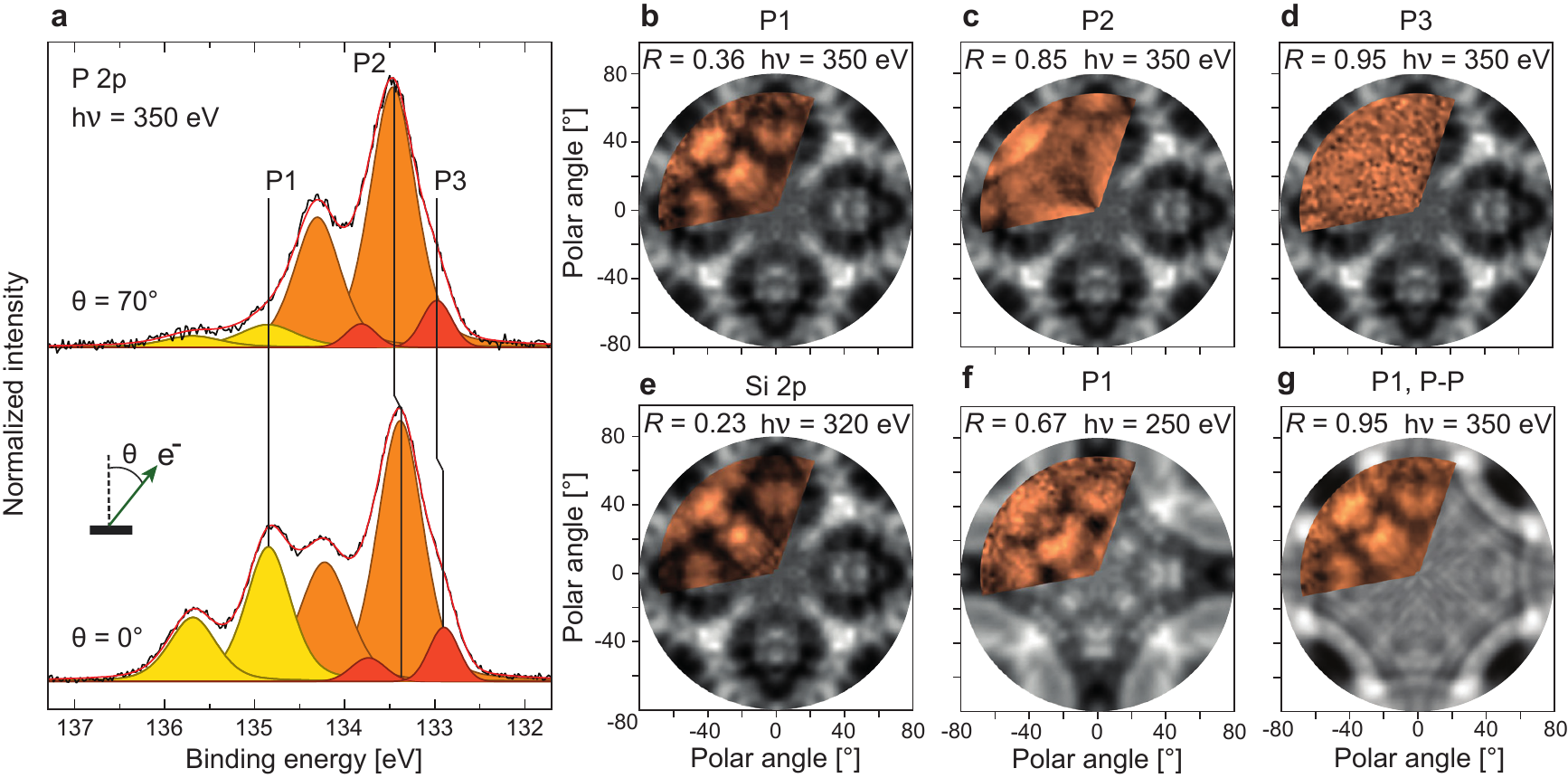}
    \caption{Angle-dependent photoelectron spectroscopy from a `double-dosed', \ce{Si}-encapsulated $\delta$-layer. \textbf{a}: XPS of the P 2p peak, measured at normal  ($\theta=0^\circ$) and grazing ($\theta=70^\circ$) emission with $h\nu=\SI{350}{\eV}$. P1 comes from P in the $\delta$-layer, and P2 and P3 from P near the \ce{Si} surface. Both spectra have been scaled to the intensity of P2. \textbf{b}-\textbf{d}: Measured (orange) and calculated (grey) XPD patterns for peaks P1-P3 from the `double-dosed' $\delta$-layer system (shown in \textbf{a}). \textbf{e}: Measured and calculated XPD from the corresponding \ce{Si}~2p core level. \textbf{f}: Measured XPD from P1 at $h\nu=\SI{250}{\eV}$, compared with XPD simulations of \ce{P}$-$\ce{Si} bonding (i.e. `substitutional doping') within the $\delta$-layer. \textbf{g}: Measured XPD of P1 at $h\nu=\SI{350}{\eV}$ (from \textbf{b}), compared with XPD simulations of P$-$P bonding (i.e. `dimerization') within the $\delta$-layer.}
    \label{fig:XPDanalysis}
\end{figure*}

The growth of $\delta$-layers has been studied and refined over the recent years, not least of all to maximize the density of P atoms within the dopant plane \cite{mckibbin2009investigating,mckibbin2014low}. The basic preparation approach involves exposing a clean Si(001) surface to a saturation coverage of \ce{PH3} gas, followed by subsequent dissociation of the gas and incorporation of P into the Si surface \cite{wang1994atomically,Wilson2004phosphine,Curson2004stm,Warschkow2005phosphine}. Refinements of the method involving multiple cycles of \ce{PH3} exposure and \ce{P} incorporation have been shown to maximize the doping density, whilst retaining a sharp confinement of the $\delta$-layer \cite{mckibbin2014low,mazzola2020sub}. In all cases, the doped surface is then overgrown with undoped silicon to encapsulate the dopant layer.

XPD, like other photoemission-based methods, is especially challenging to perform on buried atomic species because their resulting photoemission signal will be strongly attenuated by the overlayers \cite{Song2012extracting}. This problem has been addressed previously, specifically for Si:P $\delta$-layers \cite{Miwa2013direct,mazzola2014determining,cooil2017situ}. In order to demonstrate that XPD is even possible, we therefore first focus on a $\delta$-layer with a maximized dopant density (i.e. `double-dosed'), and with a minimized encapsulation layer thickness (i.e. $\approx\SI{1}{\nm}$).

Quantitative X-ray photoelectron spectroscopy (XPS) analysis (see the Methods section) of the `double-dosed' system before, during and after Si encapsulation reveals that a $0.39$ monolayer (ML) \ce{P} coverage is achieved, i.e. similar to the $0.53~\text{ML}$ reported previously \cite{mckibbin2014low}. The same analysis also reveals that $\approx90\%$ of the \ce{P} dopants remain in the $\delta$-layer after the Si overlayer growth and final annealing steps have been completed (and the additional $\approx10\%$ segregates to the surface). From our preparation we achieve an effective electron carrier density of $n=2.3\times10^{14}~\text{cm}^{-2}$ (see the Methods section for details), in line with the best-case carrier density of $n=3.6\times10^{14}~\text{cm}^{-2}$ for single-layer Si:P structures \cite{mckibbin2014low}.

The XPS signal from the phosphorous core level, after the completion of all of the growth steps, is shown as Fig.~\ref{fig:XPDanalysis}\textbf{a} at two different emission angles ($\theta$). The P~2p signal consists of three doublet components, each described by two Voigt functions with a spin-orbit splitting energy of \SI{0.84}{\eV} and an intensity ratio of p$_{3/2}$\,:\,p$_{1/2}=2\,:\,1$. The doublet labelled P1 at largest binding energy (\SI{134.85}{\eV}) corresponds to the dopants in the buried Si:P $\delta$-layer, whereas P2 (\SI{133.38}{\eV}) and P3 (\SI{132.90}{\eV}) correspond to surface phosphorus in two distinct co-ordinations \cite{Suppl_Mat}. Although $\approx90\%$ of P is present in the buried layer, the strong attenuation of the photoemission signal from buried dopants makes P1 look very weak in comparison with the un-attenuated signal (P2, P3) from trace amounts of residual surface P.

To confirm the assignment of phosphorus components from the buried $\delta$-layer and on the surface, the finite mean-free path ($\lambda$) of photoelectrons can be exploited \cite{Song2012extracting,rost2021low}. The intensity of P1 relative to P2 and P3 is seen to be strongest at normal emission ($\theta=0^\circ$) but drastically reduced at $\theta=70^\circ$.  Assuming an intensity model $I\pqty{d,\theta}\propto\text{exp}\{-d/\pqty{\lambda\cos\theta}\}$, signals from dopants at a depth $d$ beneath the surface should attenuate more rapidly with increasing $\theta$ when compared to the surface species. From this argument, we confirm that P1 is located furthest away from the surface. Investigations as a function of the photoelectron kinetic energy lead to the same conclusion \cite{Suppl_Mat}.

To determine their atomic arrangements both on and beneath the Si surface, XPD experiments of the P1-P3 components were performed. For this purpose, XPS measurements of P~2p were acquired over a large range of azimuthal ($\varphi$) and polar ($\theta$) angles, and polar plots of its intensity modulation function $\chi$ were produced alongside corresponding diffraction simulations (see the Methods section).

Since the bulk structure of Si is known \cite{kittel2018kittel}, the XPD pattern of Si~2p was also measured from one of the samples and compared to the simulated XPD pattern of the same core level as a confirmation of the methodology. The measured Si~2p XPD pattern is shown in Fig.~\ref{fig:XPDanalysis}\textbf{e} (orange) overlaid on the simulated pattern (grey), both exhibiting an apparent and similar 4-fold symmetry. Furthermore, a `reliability' factor $R=0.23$ indicates that the agreement between the two is excellent, thus confirming that the expected Si structure is well reproduced by the XPD simulation (see the Methods section for a description of the simulation optimization, and a definition of the $R$-factor).

From the high-density, `double-dosed' $\delta$-layer system, three XPD patterns of P~2p were obtained, i.e. one for each of the components P1-P3 (Figs.~\ref{fig:XPDanalysis}\textbf{b}-\textbf{d}). Notably, the measured XPD pattern of P1 (Fig.~\ref{fig:XPDanalysis}\textbf{b}) is strikingly similar to the measured XPD pattern of Si~2p at the same photoelectron kinetic energy (Fig.~\ref{fig:XPDanalysis}\textbf{e}). Matching XPD patterns from the two core levels can be expected if P and Si assume similar atomic positions: i.e, if the P1 dopant atoms replace Si atoms in the host unit cell by substitutional doping \cite{Carter2009electronic,Carter2011phosphorous}. The agreement is further supported by an achieved reliability $R=0.36$ when comparing the measured P1 XPD with an XPD simulation of subsitutional doping (orange vs. grey, Fig.~\ref{fig:XPDanalysis}\textbf{b}) \cite{Suppl_Mat}.

Contrary to the situation with P1, the XPD patterns from the surface components P2 and P3 are not expected to be well reproduced by this simulation. The measured patterns of P2 and P3 are shown in Figs.~\ref{fig:XPDanalysis}\textbf{c} and \ref{fig:XPDanalysis}\textbf{d}, respectively, overlaid on the simulated XPD from Fig.~\ref{fig:XPDanalysis}\textbf{b}. P2 shows a modulation in intensity and apparent 4-fold symmetry, but is otherwise in poor agreement with a substitutional doping model ($R=0.85$). Furthermore, P3 shows almost no structure at all, as evidenced by $R=0.96$. The achieved $R$-factors hence confirm that neither P2 nor P3 originate from bulk-substituted, sub-surface P dopants.

The XPD patterns presented so far were performed with relatively high kinetic energy photoelectrons ($E_{\text{K}}\approx\SI{220}{\eV}$), promoting forward scattering along the surface normal, and also enhancing sensitivity to the bulk structure. We also performed measurements of both Si~2p and P~2p photoelectrons with lower kinetic energy ($E_{\text{K}}\approx\SI{120}{\eV}$, Figs.~\ref{fig:XPDanalysis}\textbf{f}-\textbf{g}), i.e. intending to enhance sensitivity to the surface structure \cite{bana2018epitaxial,holt2021electronic}. To no surprise the XPD pattern was very different, and therefore the XPD simulations were further optimized to account for the apparent surface symmetry observed by surface diffraction (see the Methods section). A better agreement was achieved by means of a crude dimer model, where the surface atoms were perturbed towards a partial $2\times1$ surface reconstruction \cite{Ramstad1995theoretical,Tang1992structure}, i.e. more consistent with the observed diffraction pattern (Fig.~\ref{fig:sampleGrowth}\textbf{d}). For Si~2p at $E_{\text{K}}\approx\SI{120}{\eV}$, a perturbation of $\Delta a=\SI{0.3}{\angstrom}$ was found to give an optimal match between the measured and the simulated XPD patterns \cite{Suppl_Mat}.

Comparing the P1 XPD measured at $E_{\text{K}}\approx\SI{120}{\eV}$ with a simulation of a substitutionally doped Si:P $\delta$-layer having the same $\Delta a$ at the surface (Fig.~\ref{fig:XPDanalysis}\textbf{f}), a moderate reliability ($R=0.67$) was achieved. The higher $R$-factor found for P1 at this kinetic energy is likely related to reduced photoemission signal from -- and hence the worse statistics for, the sub-surface dopants with a shallower $\lambda$. Nonetheless, the weak reconstruction provided by the simple dimer model as described leads to a reasonable first approximation, where the main intensity modulation and symmetry of the XPD pattern is preserved.

In a simple model for \ce{PH3} dissociation on Si(001), 1-in-4 Si sites become occupied by a P atom, and 3 neighboring sites are initially occupied by H \cite{Tsukidate1999saturated,Wilson2006thermal}. This leads to the presumption that an ideal, `single-dosed' Si:P $\delta$-layer contains $25\%$ P. The local arrangement of P atoms within the $\delta$-layer has been an open debate, and multiple models have been proposed \cite{Carter2009electronic,Carter2011phosphorous}. Several of the possible arrangements include P atoms as nearest neighbors, thus leading to the suggestion of P$-$P dimers, clusters or chains \cite{Carter2011phosphorous}. When the density of P atoms on Si(001) is increased, P$-$P neighbors are expected to become increasingly common \cite{Tsukidate1999saturated}. This can potentially be problematic for Si:P derived devices, since P$-$P dimerization have been described as leading to a reduction in the overall active carrier density within the dopant layer \cite{mckibbin2014low,Keizer2015impact}.

\begin{figure}[t]
    \centering
    \includegraphics[]{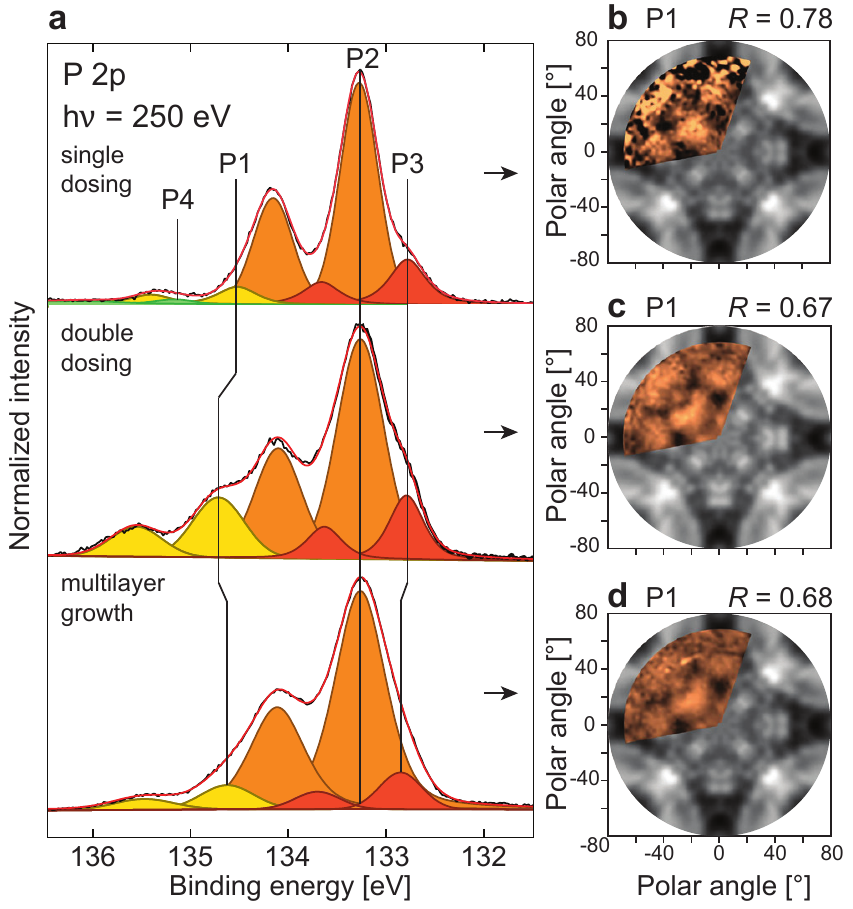}
    \caption{Comparison of the three different growth modes.\\ \textbf{a}: XPS spectra of the P~2p core level for `single-dosed', `double-dosed' and `multilayer' samples (top to bottom, respectively). The spectral intensities have been normalized to the P2 peak. \textbf{b}-\textbf{d}: Corresponding measured (orange) and simulated (grey) XPD patterns for the `single-dosed' (\textbf{b}), `double-dosed' (\textbf{c}) and `multilayer' (\textbf{d}) samples with $h\nu=\SI{250}{\eV}$.} 
    \label{fig:XPDcomparison}
\end{figure}

Our XPD study of encapsulated $\delta$-layers can offer two different insights into this matter: ($i$) We can simulate possible structures with P$-$P nearest neighbors (specifically dimers and clusters) and see if this leads to an improvement in the agreement with the experimental data, and ($ii$) we can grow a series of samples in which the dopant density within the Si is varied.

First; the measured XPD of P1 at $E_{\text{K}}\approx\SI{220}{\eV}$ is overlaid on a simulation of P$-$P dimers within the $\delta$-layer, as shown in Fig.~\ref{fig:XPDanalysis}\textbf{g} \cite{Suppl_Mat}. Both the large reliability factor $R=0.95$ and a visual comparison of the two patterns suggest that the measured and simulated XPD are poorly correlated. When measured at lower kinetic energy ($E_{\text{K}}\approx\SI{120}{\eV}$), the reliability is even worse \cite{Suppl_Mat}. We therefore infer that P$-$P dimerization has not occurred in the $\delta$-layer. Similarly, no convincing improvement was made using a cluster model \cite{Suppl_Mat}, and hence we conclude that nothing other than individual P atoms substituting in Si sites is needed to satisfactorily explain the experimental results.

Second; we prepared $\delta$-layer samples using a range of recipes in order to modify the dopant density. In addition to the sharply confined, `double-dosed', high-density recipe described above, we also prepared a lower P concentration `single-dosed'/single-layer sample with electron carrier density $n=5.1\times10^{13}~\text{cm}^{-2}$ \cite{Holt2020observation}, and a multilayer sample with 8 cycles of $\delta$-layer growth and subsequent \ce{Si} encapsulation (its $n$ similar to that of the atomically thin, `single-' and `double-dosed' $\delta$-layers \cite{mckibbin2009investigating,Keizer2015impact,Holt2020observation,mazzola2020sub}).

A comparison of the XPS and XPD structure for the three different modes of doping is shown in Fig.~\ref{fig:XPDcomparison}. The \ce{P}~2p core level spectra are shown for the three samples (`single-dosed', `double-dosed', `multilayer') in Fig.~\ref{fig:XPDcomparison}\textbf{a}. Correspondingly, the associated XPD image plots are shown in Figs.~\ref{fig:XPDcomparison}\textbf{b}-\textbf{d}. All three XPD measurements shown have been acquired at $h\nu = \SI{250}{eV}$ and follow the modulations of the P1 component from the buried dopant plane(s).

At first glance, all three XPD patterns share the same main features and symmetry as the bulk-like \ce{Si} simulation (Fig. \ref{fig:XPDanalysis}\textbf{e}). They generally have higher $R$-factors due to the reduced bulk-sensitivity at this kinetic energy ($R = 0.78$, $0.67$ and $0.68$ for single-, double- and multilayer doping, respectively). Between them, the main difference can be seen from the overall signal strength of each P1 plot, where the `single-dosed' sample has a significantly weaker intensity than the other two. Given its $\approx\SI{1}{\nm}$ \ce{Si} overlayer, the fact that P1 shows a clear modulation at all is quite impressive. The `double-dosed' sample has a stronger P1 component than the `multilayer' one, despite there being more \ce{P} dopants present in the latter. This may be a result of small differences in the overlayer thickness, and the fact that Fig.~\ref{fig:XPDcomparison}\textbf{a} shows the signal after it is normalized to the surface P2 peak: i.e. the multilayer preparation leads to an increased amount of surface \ce{P}, which makes the sub-surface \ce{P} appear relatively smaller after normalization. Nevertheless, the similarities of all three systems with bulk \ce{Si} measurements indicate that they all exhibit similar substitutional \ce{P} incorporation, and that no evidence of dimerization or clustering can be observed.

To summarize, we have first of all demonstrated that it is possible to use XPD to study the structure of dopants located beneath the surface of a semiconducting host.  Although the dopant layer is described as `high density', it is very narrow and contains a relatively small number of dopants (for example, $\approx25\%$ of an atomic layer). This makes it very challenging to study the structure with other methods. Having demonstrated the applicability of XPD, we reveal that the dopants can be accurately described as P atoms substituted into Si sites within the bulk Si crystal. This is contrary to the previous postulations of P$-$P dimer formation \cite{Carter2011phosphorous,mckibbin2014low}. Furthermore, we have used a range of sample preparation methods to create low-density, high-density, and multilayer dopant planes. We show that, in all cases, the best agreement is found by pure substitution of Si with P. Furthermore, we find no evidence to support the notion that dimerization is encouraged by increasing the dopant density or absolute dopant number.

These findings are especially important for the silicon quantum device community where Si:P $\delta$-layers are utilized as a platform. Until now the dopant structure has not been resolved, and calculations have shown that dopant ordering (such as dimerization) is an important factor in dictating the valley splitting \cite{Lee2011electronic,Carter2011phosphorous}. We therefore also conclude that XPD is an essential tool for the development and optimization of quantum device architectures based on sub-surface dopant assemblies -- such as the much prized $\delta$-layer platform.

\section{Methods}
\subsection{Sample Growth}
Surfaces of $n$-type Si(001) with negligible surface oxide on them were prepared in-vacuum by short cycles of high-temperature annealing to \SI{1200}{\celsius} (measured with pyrometer, $\epsilon=0.79$). The clean surfaces revealed a $(2\times1)$ reconstruction when investigated using low-energy electron diffraction (LEED), as shown in Fig.~\ref{fig:sampleGrowth}\textbf{b}. Next, the surfaces were exposed to $1.125$~Langmuirs (L) of gaseous \ce{PH3} (partial pressure $5\times10^{-9}$~mbar for 5 mins) and subsequently annealed to \SI{550}{\celsius} to dissociate the \ce{PH3} and incorporate \ce{P} into the Si surface \cite{Curson2004stm,mckibbin2009investigating}. For the `double-dosed' samples, dosing of \ce{PH3} and subsequent annealing were repeated twice \cite{mckibbin2014low}. For the multilayer samples, 8 cycles of \ce{PH3} dosing, annealing and subsequent deposition of 1 atomic Si layer were performed \cite{Keizer2015impact}. Finally, all dopants were encapsulated by $\approx\SI{1}{\nm}$ Si and given a short, post-deposition anneal to \SI{350}{\celsius} for a few seconds. This triggered a $(2\times1)$ phase re-ordering of the Si surface (Figs.~\ref{fig:sampleGrowth}\textbf{c}, \ref{fig:sampleGrowth}\textbf{d}).

\subsection{Photoemission Measurements}
High-resolution X-ray photoelectron spectroscopy (XPS) measurements of the \ce{Si}~2p and \ce{P}~2p core levels were performed, throughout the preparation of the `single-dosed', `double-dosed', and `multilayer' samples. For each finished structure, the same core levels were subsequently measured using X-ray photoelectron diffraction (XPD). All photoemission measurements were performed at the SuperESCA endstation of Elettra Synchrotron in Trieste, Italy. All spectra were collected at room temperature ($T\approx\SI{300}{\kelvin}$), using a SPECS Phoibos electron energy analyzer equipped with a homemade delay-line detector. The overall energy resolution was $\Delta E<\SI{50}{\milli\eV}$ for all the measurements. The photoexcitation energies $h\nu$ were calibrated from the kinetic energy difference of Si~2p peaks that were collected using first- and second-order light from the monochromator.

\subsubsection{XPS Analysis}
\begin{figure}[t]
    \centering
    \includegraphics[]{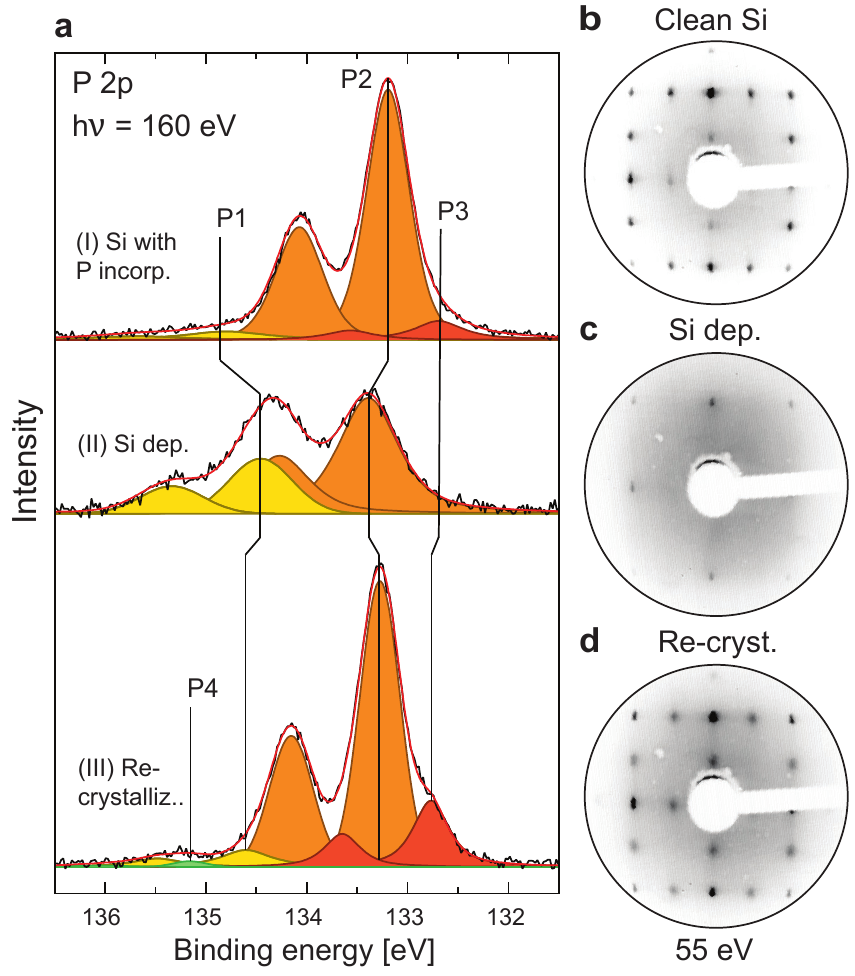}
    \caption{Epitaxial growth of a `single-dosed' Si:P $\delta$-layer.\\
    \textbf{a}: The development of the P~2p core level measured at $h\nu=\SI{160}{\eV}$, upon (I) \ce{PH3} decomposition and dopant incorporation, (II) encapsulation/Si overlayer deposition, and (III) re-crystallization of the Si overlayer. \textbf{b}-\textbf{d}: Surface diffraction (LEED) patterns of the `single-dosed' $\delta$-layer system (\textbf{b}) before doping, (\textbf{c}) after doping and Si encapsulation, and (\textbf{d}) after re-crystallization of the Si overlayer.}
    \label{fig:sampleGrowth}
\end{figure}

The characterization of the `single-dosed' $\delta$-layer system is summarized in Fig.~\ref{fig:sampleGrowth}. Starting with a clean, (2$\times$1)-reconstructed Si(001) surface (Fig.~\ref{fig:sampleGrowth}\textbf{b}) at room temperature, gaseous \ce{PH3} was adsorbed and partially dissociated \cite{Lin1999thermal,Lin2000interaction,Wilson2004phosphine,Warschkow2005phosphine}. An overall P surface coverage of 0.17 monolayer (ML) was achieved, as estimated from a simple two-layer attenuation model \cite{rost2021low}.

Upon annealing, three different doublet components became visible from the P~2p core level (Fig.~\ref{fig:sampleGrowth}\textbf{a}, I). Each doublet can be described by a Voigt line shape with a spin-orbit energy splitting of \SI{0.84}{\eV} and an intensity ratio of $\text{p}_{3/2}:\text{p}_{1/2}=2:1$. According to the re-interpretation by Wilson \emph{et al.} \cite{Wilson2006thermal}, the main doublet P2 at binding energy $E_{\text{P2}}=\SI{133.20}{\eV}$ should be from surface-incorporated \ce{P} species. The other two doublets P1 and P3 appear at the relative energies $+\SI{1.60}{\eV}$ and $-\SI{0.51}{\eV}$, respectively. These have been interpreted as different species of surface \ce{P} with variations in their local atomic environment \cite{Suppl_Mat}.

During Si encapsulation (Fig.~\ref{fig:sampleGrowth}a, II), the surface P2 intensity reduced linearly vs. time, as opposed to the negative exponential decay expected from a simple attenuation model \cite{rost2021low}. Additionally, P1 grew in absolute terms. Since a finite amount of P was present in the system, and P2 was assumed to be located at the sample surface at all times (i.e. not attenuated by the addition of Si), the loss of P2 intensity was used to estimate the amount of P1 formed as $-\Delta\text{P2}=+\Delta\text{P1}$.

A final anneal to \SI{350}{\celsius} triggered a re-crystallization of the overlayer, as evidenced by the relatively sharp LEED spots and the recurrence of the 2$\times$1 reconstruction (Figs.~\ref{fig:sampleGrowth}\textbf{c}, \ref{fig:sampleGrowth}\textbf{d}). This final anneal also promoted segregation of \ce{P} to the surface, and hence the intensity of the surface components (P2 and P3) increased in the XPS (Fig.~\ref{fig:sampleGrowth}\textbf{a}, II-III). Additionally, a very weak component P4 appeared only in the `single-dosed case', with a disordered spatial structure as determined from its XPD pattern (see the Supplementary note \cite{Suppl_Mat}).

By tracing the development the P1 intensity in Fig.~\ref{fig:sampleGrowth} (steps I-III) as described, and comparing its final ML coverage to the atomic packing density of the Si(001) plane, the effective electron carrier density $n$ was estimated. From `single-dosing' (Fig.~\ref{fig:sampleGrowth}\textbf{a}), the $n$ achieved was estimated to be $n=5.1\times10^{13}~\text{cm}^{-2}$ in the buried $\delta$-layer. This is well above the metal-to-insulator transition \cite{kravchenko2003metal}, and roughly consistent with previous reports \cite{Lin1999thermal,mckibbin2009investigating,Holt2020observation}.

\subsubsection{XPD Analysis}
XPD patterns from each finished sample were produced by measuring the \ce{Si}~2p and \ce{P}~2p core levels -- therein including the sub-components of the buried $\delta$-layer, over a wide azimuthal sector ($\varphi=0-130^\circ$), and from grazing ($\theta=70^\circ$) to normal emission ($\theta=0^\circ$). Each measured spectrum (851 per XPD pattern) was fitted with symmetric Voigt functions to deconvolve the various sub-components present. The intensity $I\pqty{\theta,\varphi}$ of each fitted sub-component was then used to produce polar plots of their modulation functions $\chi$ (commonly referred to as `stereographic projections' \cite{Bignardi2019growth}), defined as
\begin{equation}
    \chi = \dfrac{I\pqty{\theta,\varphi}-I_{0}\pqty{\theta}}{I_{0}\pqty{\theta}},
    \label{eq:ModulationFn}
\end{equation} 
where $I_{0}\pqty{\theta}$ is the average intensity for a given $\theta$ across all the azimuthal ($\varphi$) scans \cite{holt2021electronic}.

\subsection{Determining the $\delta$-Layer Structure}
Simulated diffraction patterns from different dopant positions were obtained using the `Electron Diffraction in Atomic Clusters' (EDAC) package \cite{Garcia2001multiple}. The degree of agreement between each measured and simulated diffraction pattern was quantified by a `reliability' factor $R$:
\begin{equation}\label{eq:Rfactor}
	R = \dfrac{\sum_{i}\pqty{\chi_{\text{sim},i}-\chi_{\text{exp},i}}^{2}}{\sum_{i}\pqty{\chi_{\text{sim},i}^{2}+\chi_{\text{exp},i}^{2}}},
\end{equation}
where $\chi_{\text{exp},i}$ and $\chi_{\text{sim},i}$ correspond to the experimental and simulated intensity modulation functions, respectively. The sum index $i$ runs over all available data points at the different angles measured. The lower the $R$, the better the agreement between the experiment and the atomic model ($R=0$ corresponds to a complete agreement; $R=1$ means no correlation; $R=2$ signifies anti-correlation \cite{Woodruff2007adsorbate}). The best understanding of the atomic arrangement was determined by minimizing $R$ upon iterative adjustments of the simulated XPD, with subsequent comparison to the experimental XPD, until an optimum fit between the two was reached.

\section{Data Availability}
The data underpinning the findings presented in this publication can be made available from the corresponding author upon reasonable request.

\section{Acknowledgements}
This work was partly supported by the Research Council of Norway (RCN) through project numbers 324183, 315330, and 262633. Additional financial support was received from CALIPSOplus, under Grant Agreement 730872 from the EU Framework Programme for Research and Innovation HORIZON 2020. We acknowledge Elettra Sincrotrone Trieste for providing access to its synchrotron radiation facilities and for all technical assistance. We would also like to thank Ph. Hofmann, J. A. Miwa, M. Bianchi, F. Mazzola, S. P. Cooil, A. J. Holt, and J. Bakkelund for fruitful discussions.

\section{Author Contributions}
H.I.R., F.S.S., A.C.\AA., and J.W.W. measured the XPS and LEED data, which was in turn analyzed by H.I.R. and A.C.\AA. The XPD simulations were performed by E.T. with input from all co-authors. H.I.R., F.S.S., A.C.\AA.,  J.W.W.  E.T., P.L. and S.L. operated the SuperESCA endstation and performed the XPD measurements. The project was conceived and led by J.W.W. The manuscript was written by H.I.R., F.S.S., A.C.\AA., and J.W.W. with contributions from all the authors.

% Bibliography
%\bibliography{delta}% Produces the bibliography via BibTeX.

%apsrev4-2.bst 2019-01-14 (MD) hand-edited version of apsrev4-1.bst
%Control: key (0)
%Control: author (8) initials jnrlst
%Control: editor formatted (1) identically to author
%Control: production of article title (0) allowed
%Control: page (0) single
%Control: year (1) truncated
%Control: production of eprint (0) enabled
%

\end{document}

\title{Supplementary Note:\\Probing the Atomic Arrangement of Sub-Surface Dopants in a Silicon Quantum Device Platform}%

% Authors
\author{H{\aa}kon I. R{\o}st}
\affiliation{Center for Quantum Spintronics, Department of Physics, Norwegian University of Science and Technology (NTNU), NO-7491 Trondheim, Norway.}
\affiliation{Department of Physics and Technology, University of Bergen (UiB), All\'egaten 55, 5007 Bergen, Norway.}

\author{Ezequiel Tosi}
\affiliation{Elettra - Sincrotrone Trieste, s.s. 14 - km.163,5 in Area Science Park, Basovizza, Trieste, 34149, Italy.}
\affiliation{Instituto de Ciencia de Materiales de Madrid (ICMM-CSIC) C/ Sor Juana Inés de la Cruz 3, 28049, Madrid, Spain.}

\author{Frode S. Strand}
\author{Anna Cecilie {\AA}sland}
\affiliation{Center for Quantum Spintronics, Department of Physics, Norwegian University of Science and Technology (NTNU), NO-7491 Trondheim, Norway.}

\author{Paolo Lacovig}
\author{Silvano Lizzit}
\affiliation{Elettra - Sincrotrone Trieste, s.s. 14 - km.163,5 in Area Science Park, Basovizza, Trieste, 34149, Italy.}

\author{Justin W. Wells}
\email[Corresponding author: ]{j.w.wells@fys.uio.no}
\affiliation{Center for Quantum Spintronics, Department of Physics, Norwegian University of Science and Technology (NTNU), NO-7491 Trondheim, Norway.}
\affiliation{Centre for Materials Science and Nanotechnology, University of Oslo (UiO), Oslo 0318, Norway.}

\maketitle
\section*{Surface $\text{PH}_{3}$ Adsorption and $\text{P}$ Dopant Incorporation}
At room temperature ($T\approx\SI{300}{\kelvin}$), gaseous phosphine is known to dissociate and bond with the \ce{Si}$-$\ce{Si} dimers on Si(001), following a pathway of three successive \ce{H} deficient products with different reaction rates \cite{Wilson2006thermal}. In a manner of minutes, the adsorbed gas rapidly dissociates from $\ce{PH3}\rightarrow\ce{PH2}+\ce{H}$, then $\ce{PH2}\rightarrow\ce{PH}+\ce{H}$, and finally $\ce{PH}\rightarrow\ce{P}+\ce{H}$ \cite{Wilson2004phosphine,Warschkow2005phosphine}. At temperatures $T>\SI{650}{\kelvin}$, the end product is seen to release $\ce{H2}$ gas and incorporate \ce{P} into the surface by substitution of \ce{Si} atoms \cite{wang1994atomically,Curson2004stm,mckibbin2009investigating}.

\begin{figure}[t]
    \centering
    \includegraphics[]{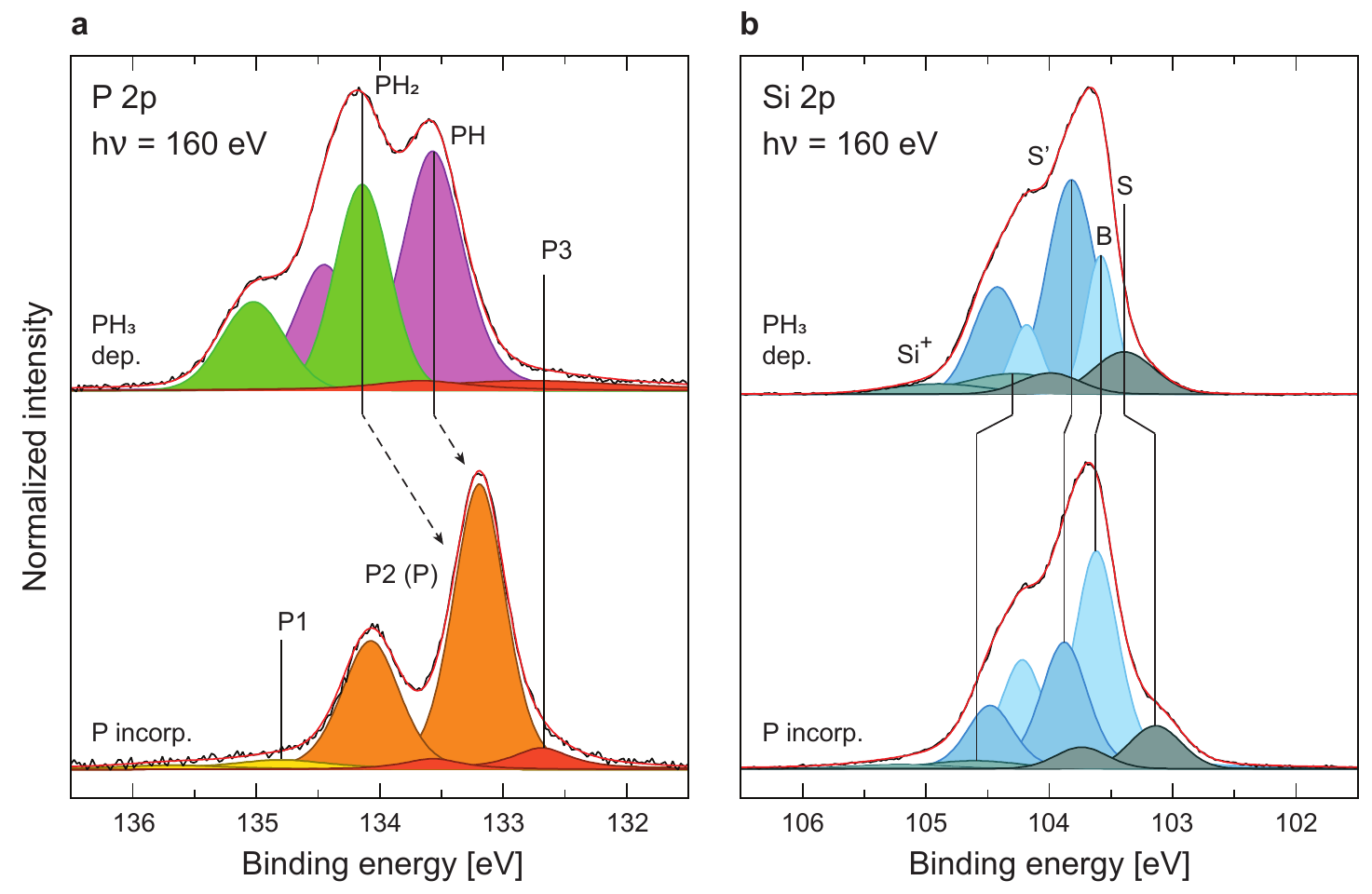}
    \caption{XPS of \ce{P}~2p and \ce{Si}~2p at \ce{PH3} adsorbtion and \ce{P} dopant incorporation. \textbf{a}: Components \ce{PH2}, \ce{PH} and P3 are dissociation products of adsorbed \ce{PH3}, while P2 and P1 are P dopants in and beneath the Si(100) surface, respectively \cite{Wilson2004phosphine,Lin1999thermal}. \textbf{b}: Component B stems from bulk Si while S' and S from surface \ce{Si} dimers and second layer atoms, respectively \cite{Lin1991dimer}. The tiny $\text{Si}^{+}$ is from remnants of intermediary oxide at the surface \cite{hollinger1984probing}. All spectra have been normalized to the maximum peak intensity of each core level.}
    \label{fig:figS1}
\end{figure}

An X-ray photoelectron spectroscopy (XPS) study of room temperature adsorption and dissociation of phosphine gas onto Si(001) and subsequent, thermally promoted surface incorporation of \ce{P} atoms is shown in Fig.~\ref{fig:figS1}. Upon adsorption, two leading doublets can be distinguished from the \ce{P}~2p core level in the range $133$-$\SI{135}{\eV}$, along with a faint broad feature at lower binding energies. Based on the previous XPS studies by Lin \emph{et al.} and their re-interpretation by Wilson \emph{et al.}, the two main doublets can be assigned to surface-bound \ce{PH} (\SI{133.57}{\eV}) and \ce{PH2} (\SI{134.15}{\eV}) \cite{Lin1999thermal,Wilson2006thermal}. The origin of the final component P3 is unclear, but it is assumed to be \ce{PH_{x}} at a different, less energetically favorable binding site (see Fig.~2 of Ref.~\cite{Warschkow2005phosphine}). The \ce{Si}~2p core level has four doublets: three are known bulk- (B, \SI{103.58}{\eV}) and surface-related (S', \SI{103.82}{\eV}; S, \SI{103.40}{\eV}) components \cite{Lin1991dimer} and the final one matches with $\ce{Si}^{+}$ from a minuscule amount of intermediary \ce{Si2O} oxide (\SI{104.29}{\eV}) \cite{hollinger1984probing}.

Upon thermal activation to $T=\SI{820}{\kelvin}$ the \ce{PH} and \ce{PH2} signals are depleted, and a new leading feature P2 appears at \SI{133.20}{\eV}. A similar change was observed by Lin \emph{et al.} (Fig. 3b, Ref.~\cite{Lin1999thermal}) and can be attributed to the incorporation of P atoms into the Si surface \cite{Wilson2006thermal}. P3 is still present but in higher quantity, and a new feature P1 appears at \SI{134.80}{\eV} far away from any of the pre-existing dissociation components. This latter feature can be assigned to P dopants beneath the \ce{Si} surface, as made evident from the discussions in the main paper and the energy-dependent XPS in the next section.

\section*{Probing Si-encapsulated P dopants with increasing $\lambda$}
In photoemission experiments, the kinetic energy $\varepsilon$ dependence of the inelastic mean-free path $\lambda\pqty{\varepsilon}$ for outgoing photoelectrons can be used to determine the layer structure of a system. On average, $\lambda\pqty{\varepsilon}\propto\varepsilon^{-2}$ for low kinetic energies ($\varepsilon\leq\SI{30}{\eV}$), whilst for higher kinetic energies ($\varepsilon>\SI{75}{\eV}$) then $\lambda\pqty{\varepsilon}\propto\varepsilon^{1/2}$ \cite{seah1979quantitative}. Changing $\varepsilon$ will thus change the average effective escape depth of photoelectrons from beneath the surface of a material.

A simple, yet effective attenuation model for photoelectron intensity can be constructed from the Beer-Lambert law as
\begin{equation}
     I\pqty{d,\theta}\propto I_{0}\exp{-d/\bqty{\lambda\pqty{\varepsilon}\cos\theta}},
\end{equation}
where $d$ is the probing depth in the direction orthogonal to the surface, $\theta$ the emission angle relative to the surface normal, and $I_{0}$ is the signal intensity from the surface ($d=0$) \cite{Song2012extracting}. According to the model, signals from features at larger $d$ should attenuate exponentially, and also roll off more rapidly with $\theta$ than signals from closer to the surface. Furthermore, the attenuation can be controlled by changing the ratio $d/\lambda\pqty{\varepsilon}$. For instance, the detectable signal from features that are spatially localized along $d$, e.g. an atomically thin, encapsulated $\delta$-layer, can be either enhanced or diminished by varying $\lambda(\varepsilon)$.

\begin{figure}[t]
    \centering
    \includegraphics[]{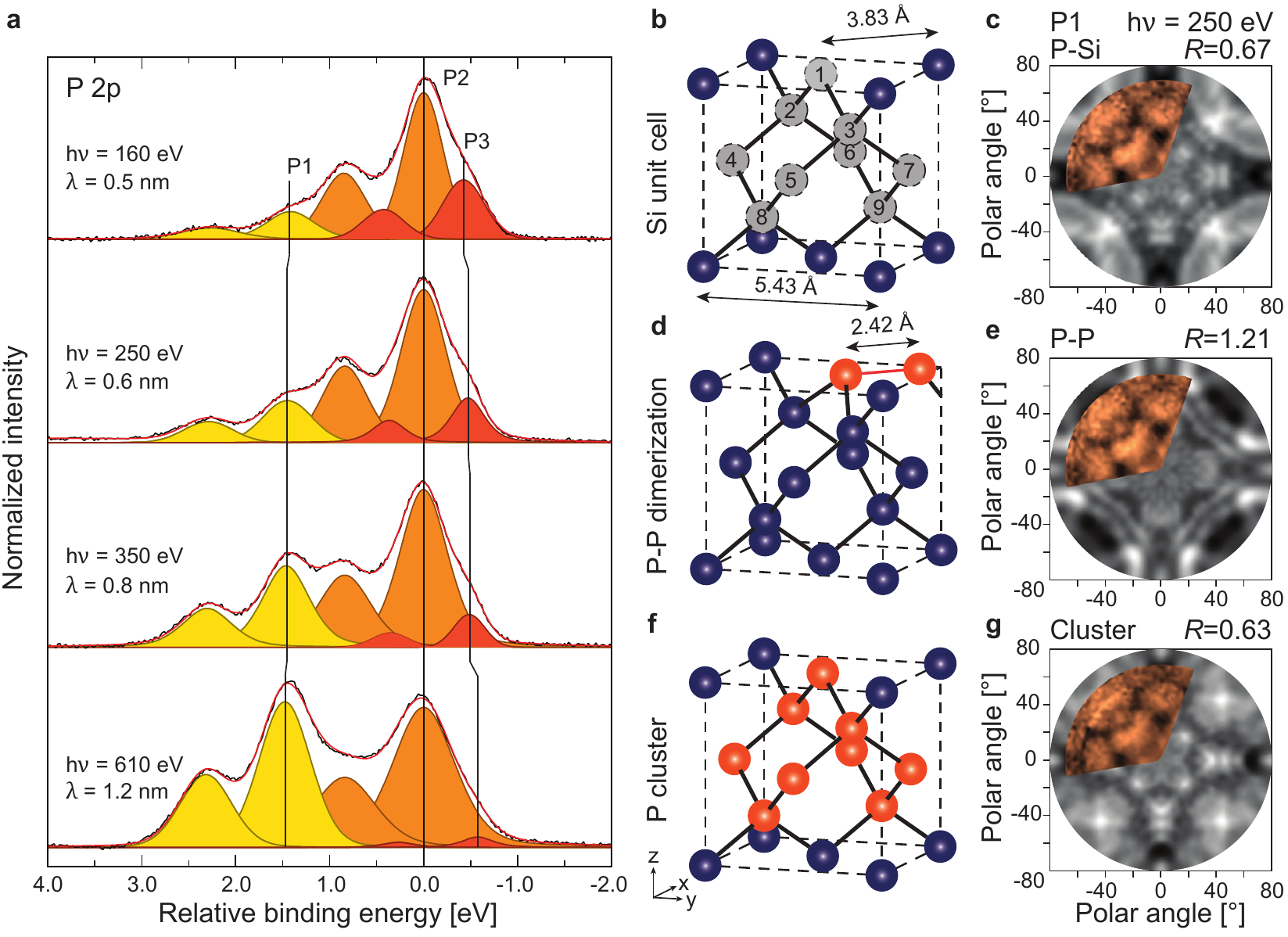}
    \caption{Energy-dependent P~2p core level analysis and modeling of P dopant atom placement. \textbf{a}: The measured \ce{P}~2p signal from a `double-dosed' $\delta$-layer sample, as a function of increasing photoelectron kinetic energy $\varepsilon$ and inelastic mean-free path $\lambda\pqty{\varepsilon}$. The core levels have been plotted on a binding energy scale relative to the P2 feature ($E_{\text{B}}=\SI{133.20}{\eV}$). All spectra have been normalized to the measured peak maximum intensity. \textbf{b}--\textbf{g}: Lattice models for P atom placements in $\delta$-doped Si and their corresponding, simulated XPD patterns. In the Si fcc unit cell \cite{Ramstad1995theoretical}, there are 9 inequivalent atomic positions. For substitutional doping (\textbf{b}, \textbf{c}), either site can be occupied by a P atom and will have Si atoms as its neighbors. For in-plane P$-$P dimerization (\textbf{d}, \textbf{e}), Si atoms on the $(001)$ face are replaced, and the newfound P atoms relaxed towards each other to form bonds. In a `cluster' model (\textbf{f}, \textbf{g}), all 9 inequivalent Si atoms get replaced simultaneously.}
    \label{fig:figS2}
    \vspace{-0.4cm}
\end{figure}

The measured \ce{P}~2p signal from a `double-dosed' $\delta$-layer is shown in Fig.~\ref{fig:figS2}\textbf{a} as a function of increasing $\lambda\pqty{\varepsilon}$. Notably, at larger $\lambda\pqty{\varepsilon}$ the intensities of P1 and P3 relative to P2 increase and decrease, respectively. As discussed previously, P2 can be assigned to incorporated P dopants near the surface. The relative increase in P1, therefore, suggests that as $\lambda\pqty{\varepsilon}$ increases, more P1 signal is being added into the increasingly thicker `slab' of excitation volume beneath the surface. This matches with the P1 signal originating from dopants in the $\delta$-layer that are buried beneath the surface. Additionally, the relatively decreasing P3 signal suggests that this specie is local to the Si surface, perhaps even more so than P2.

\section*{Atomic Models for P Dopant Placement in Si}
Different XPD patterns expected from different P dopant configurations were simulated to compare with the measured X-ray photoelectron diffraction (XPD) of the P~2p core level (Figs.~1-2, main text), using the `Electron Diffraction in Atomic Clusters' (EDAC) package \cite{Garcia2001multiple}. Sketches of the atomic models used for the dopant placement, and their corresponding simulated XPD patterns compared to the measured P~2p signal, are shown in Figs.~\ref{fig:figS2}\textbf{b}-\textbf{g}.

To model substitutional doping, Si atoms in the fcc `diamond'-like unit cell were replaced by P atoms one by one, i.e. so that each possible P dopant position would neighbor with Si atoms. The resulting XPD pattern for each inequivalent electron emitter (i.e. P atom) position was then simulated. Finally, all contributions (9 in total) were added together to form the resultant, total XPD pattern. In the case of P$-$P dimerization, adjacent Si atoms in the Si$(001)$ plane were replaced by P atoms and subsequently relaxed towards each other to form P$-$P bonds along the $[110]$ direction. An optimum bond length of  $d=\SI{2.42}{\angstrom}$ was found, i.e. slightly longer than the bulk Si$-$Si bond length (\SI{2.35}{\angstrom}) \cite{Tang1992structure}, and an intermediate between other reported P$-$P bond lengths \cite{cataldo2001formation}. For a `cluster' model, all 9 inequivalent Si atoms in the unit cell were simultaneously replaced by P atoms, retaining the same nearest neighbor bond length as that of bulk Si.

At $h\nu=\SI{250}{\eV}$ ($\varepsilon\approx\SI{115}{\eV}$) the measured XPD signal from the P dopants (P1) achieved a moderate agreement with the substitutional and cluster-like doping models ($R=0.67$ and $R=0.63$, respectively). The P$-$P dimer model ($R=1.21$) showed no correlation with the measurements and was therefore ruled out completely. Having already established the confined placement of the P dopants beneath the Si surface (Fig.~\ref{fig:figS2}\textbf{a}, and Fig.~1\textbf{a} in the main text), high-concentration cluster-like doping could also be ruled out. Hence substitutional was the only likely scenario.

\section*{Simulating XPD with the Correct Si Surface Structure}
\begin{figure}[]
    \centering
    \includegraphics[]{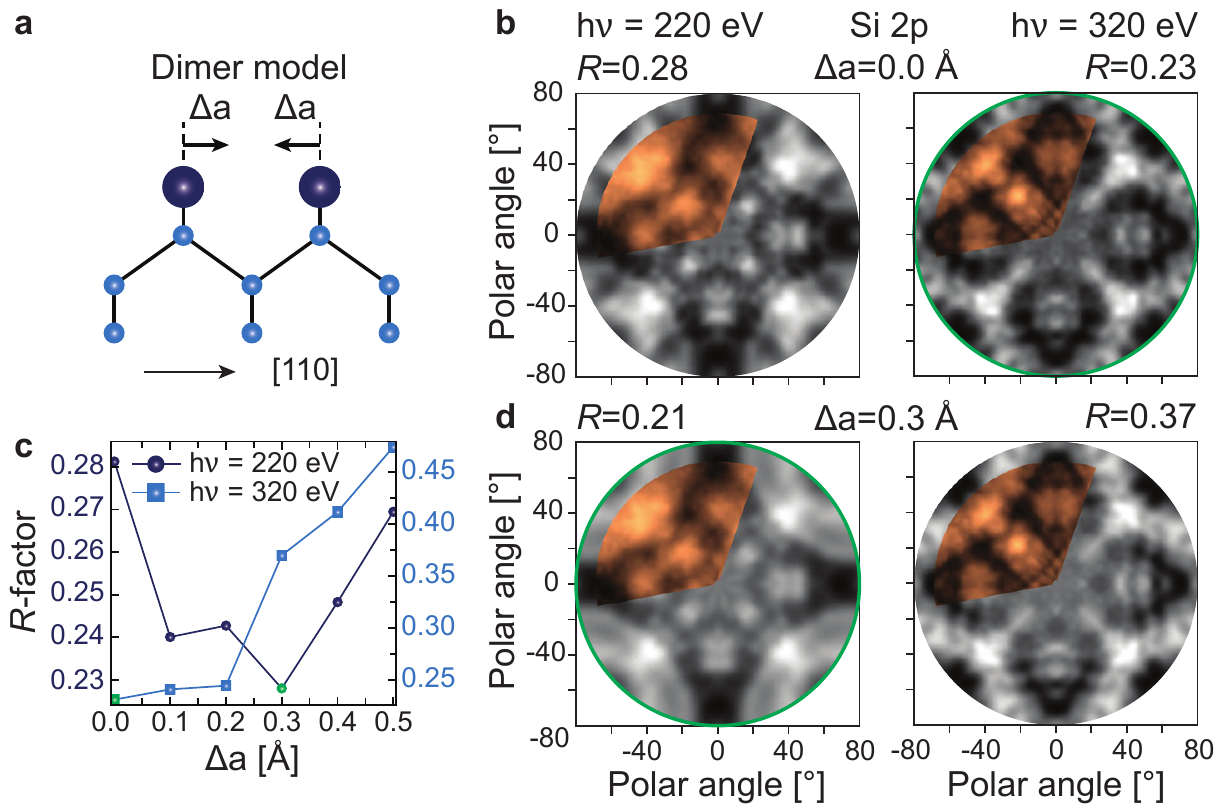}
    \caption{Si surface dimerization and kinetic energy-dependent XPD measurements of Si:P $\delta$-layers. \textbf{a}: A side view, ball-and-stick model of a dimerizing Si(001) surface along the $[110]$ direction. \textbf{b}--\textbf{d}: Reliability ($R$) factor optimization for Si~2p photoelectrons excited with different photon energies $h\nu$, as a function of Si surface atom displacement $\Delta a$ towards dimerization (\textbf{c}). Example comparisons of the measured (orange) and simulated (grey) XPD patterns, shown for two different displacements $\Delta a$ with the best fit at each $h\nu$ circled in green (\textbf{b}, \textbf{d}). The measurements shown with $h\nu=\SI{220}{\eV}$ and $h\nu=\SI{320}{\eV}$ are from a `single-dosed' and `double-dosed' system, respectively.}
    \label{fig:figS3}
\end{figure}

For comparison with the more surface-sensitive XPD measurements ($h\nu=220-\SI{250}{\eV}$), the Si lattice model was adjusted to account for the observed ($2\times1$) reconstruction at the surface of the encapsulation layer (see Fig.~3\textbf{d} in the main text). Using a crude model that ignores any relaxation between the topmost atomic layers, the \ce{Si} atoms in the surface layer were perturbed towards each other by a distance $\Delta a~[\text{Å}]$ (Fig.~\ref{fig:figS3}\textbf{a}), yielding an optimal reliability factor $R=0.23$ at $\Delta a=\SI{0.3}{\angstrom}$ (Figs.~\ref{fig:figS3}\textbf{c}, \ref{fig:figS3}\textbf{d}). Note that this displacement is too small for a proper Si$-$Si dimerization to occur \cite{Tang1992structure,Ramstad1995theoretical}. Therefore, it merely serves as a first step toward a more accurate surface model. In contrast, a negligible $\Delta a$ was preferred to minimize $R$ when comparing with measurements of photoelectrons with higher kinetic energies $\varepsilon$, i.e. when using $h\nu=320-\SI{350}{\eV}$ (Figs.~\ref{fig:figS3}\textbf{b}, \ref{fig:figS3}\textbf{c}). This can be explained by the expected reduction in electron scattering along other directions than the surface normal as $\varepsilon$ increases, and also the increased bulk sensitivity achieved with an increasing inelastic mean-free path $\lambda(\varepsilon)$ \cite{seah1979quantitative,Woodruff2007adsorbate}. 

\section*{XPD of `single-dosed' P~2p sub-components}
\begin{figure}[]
    \centering
    \includegraphics[]{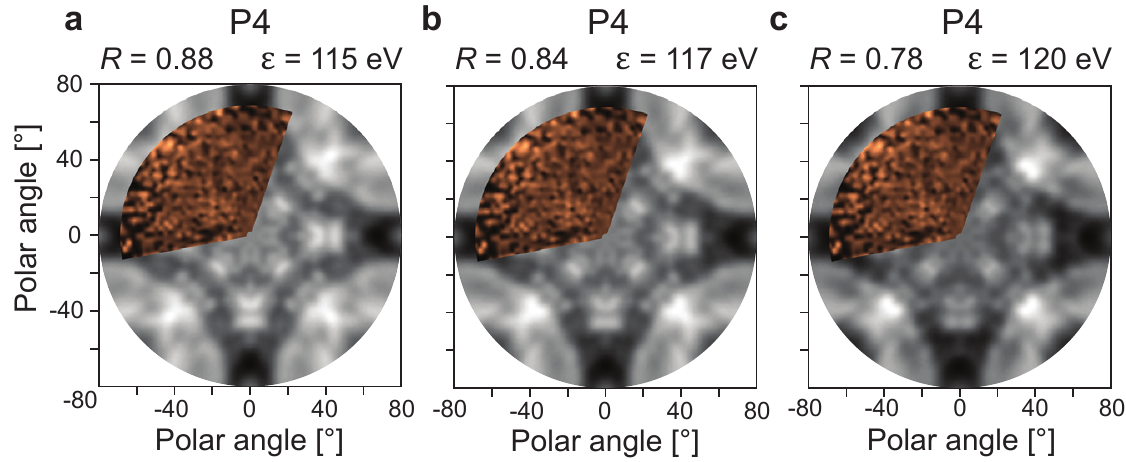}
    \caption{Kinetic energy-dependent reliability ($R$) factor of component P4. Comparison of the measured and simulated XPD patterns using the optimized structural model at $h\nu=\SI{250}{\eV}$. The $R$ at $\varepsilon=\SI{115}{\eV}$ (\textbf{a}), which is closest to the true kinetic energy of the signal, is worse than the $R$ several eV away from the peak position (\textbf{b}, \textbf{c}).}
    \label{fig:figS4}
\end{figure}

In Fig.~\ref{fig:figS4}, XPD patterns of the P4 sub-component of the P~2p signal that was measured from a `single-dosed' $\delta$-layer sample are shown. With the `surface-sensitive' photoexcitation energy used ($h\nu=\SI{250}{\eV}$), the P4 was ejected with a kinetic energy $\varepsilon\approx\SI{115}{\eV}$. As shown in Fig.~\ref{fig:figS4}\textbf{a}, only a faint ordering is visible that is vaguely reminiscent of the one observed from P1 and bulk Si. When compared with the corresponding, optimized XPD simulation of Si at a similar $\varepsilon$ (see the previous Section), a reliability factor of $R=0.88$ is achieved. In other words, there is almost no agreement between the measurements and the calculations. Interestingly, the $R$-factor improves at larger $\varepsilon$ away from the peak, i.e. when moving towards the energy of the P1 signal originating from the buried $\delta$-layer (Figs.~\ref{fig:figS4}\textbf{b}, \textbf{c}). The reason for this is unclear, but may perhaps suggest that the species P1 and P4 have some similarities in their atomic arrangements. Ultimately, no strong evidence for atomic ordering of the P4 component can be observed.

\newpage

% Bibliography
%\bibliography{deltaSupp}% Produces the bibliography via BibTeX.

%apsrev4-2.bst 2019-01-14 (MD) hand-edited version of apsrev4-1.bst
%Control: key (0)
%Control: author (8) initials jnrlst
%Control: editor formatted (1) identically to author
%Control: production of article title (0) allowed
%Control: page (0) single
%Control: year (1) truncated
%Control: production of eprint (0) enabled
%